\documentclass[12pt]{article}

\begin{document}

\begin{titlepage}

\topmargin=3.5cm

\textwidth=13.5cm

\centerline{\Large \bf Equivariant Localization for}

\vspace{0.4cm}

\centerline{\Large \bf Supersymmetric Quantum Mechanics}

\vspace{1.0cm}

\centerline{\large \bf Levent Akant\footnote{E-mail:
lakant@yeditepe.edu.tr}}

\vspace{0.5cm}

\centerline{ \textit{Yeditepe University, Department of Physics}}

\centerline{\textit{Inonu Mah. Kayisdagi Cad. Kadikoy, Istanbul
34755}}

\vspace{1.5cm}

We apply equivariant localization to supersymmetric quantum
mechanics and show that the partition function localizes on the
instantons of the theory. Our construction of equivariant
cohomology for SUSY quantum mechanics is different than the ones
that already exist in the literature. A hidden bosonic symmetry is
made explicit and the supersymmetry is extended. New bosonic
symmetry is the square of the new fermionic symmetry. The $D$ term
is now the parameter of the bosonic symmetry. This construction
provides us with an equivariant complex together with a Cartan
differential and makes the use of localization principle possible.

\end{titlepage}

\baselineskip=0.6cm

\abovedisplayskip=0.4cm

\belowdisplayskip=0.4cm

\abovedisplayshortskip=0.3cm

\belowdisplayshortskip=0.3cm

\jot=0.35cm

\section{Introduction}
The aim of this article is to apply equivariant localization to
SUSY quantum mechanics and show that the partition function of the
theory localizes on the instantons of the action. This is a well
known result that can also be derived by using arguments based on
Nicolai map \cite{Nicolai}, \cite{BT1}. The advantage of
localization principle is that it is a very general result, almost
as general as Stoke's theorem of ordinary de Rham theory, which
can easily be applied whenever the right setting, namely the
equivariant cohomology, is present. There are important instances
in quantum field theory where such a setting is available.
\cite{Atiyah, Bismut, Witten1, Witten2, BT2, Morozov}. We will
show that SUSY quantum mechanics is another example which
accommodates equivariant cohomology, and use localization
principle to derive the above mentioned result. Our work will also
make the symplectic structure of the theory more explicit.

The construction presented in this article is different than the
one used by Morozov, Niemi and Palo in their work on SUSY theories
\cite{Morozov} and the one used by Cordes, Moore and Ramgoolam in
\cite{Moore}. The main difference is that in our case the BRST
like operator which gives rise to equivariant localization is not
nilpotent; instead it squares to a hidden bosonic symmetry of the
theory. So the BRST operator is nilpotent only on the gauge
invariant observables of the theory. With this, our construction
becomes a special abelian case of the general construction given
by Witten in \cite{Witten1}.

Equivariant cohomology is a generalization of de Rham theory to
manifolds with group actions. If a group $G$ acts freely on a
manifold $M$ then equivariant cohomology is defined to be the de
Rham cohomology of the quotient manifold $M/G$. If the action is
not free it is still possible to define equivariant cohomology as
cohomology of certain complexes associated with $M$ and $G$. Two
commonly used complexes are the Weil complex and the Cartan
complex. Resulting cohomologies are equivalent and reduce to de
Rham cohomology of $M/G$ in the case of free actions. In this
article we will use the Cartan model of equivariant cohomology.
For reviews of equivariant cohomology see \cite{Atiyah-Bott, GS,
Szabo, BT3, Moore}.

The relevance of equivariant cohomology to quantum field theory is
that the field content of theories with BRST symmetry fit into an
equivariant complex (Cartan complex). Therefore one can hope to
apply many interesting and useful results of equivariant
cohomology to the analysis of these quantum field theories. One
such result is localization.

In de Rham theory one can integrate differential forms. The
theorems of classical integral calculus such as Stoke's theorem
and divergence theorem are naturally unified in exterior calculus
and this provides us with great technical prowess and deep
geometric insights. In equivariant theory one can integrate
equivariant forms. The most interesting result in equivariant
integration is localization. It means that the integral of a
closed, but otherwise arbitrary, equivariant form over a large
region is equal to the integral of the same form over a much
smaller subregion. In quantum field theory this means that only a
subset of field configurations contribute to the path integral. A
particularly useful special case of localization arises when the
manifold is symplectic and the action of the group is Hamiltonian.
All the known applications of equivariant localization to field
theory, including this one, involve infinite dimensional analogs
of this case.

Localization was first proved by Duistermaat and Heckman in the
case of torus actions on symplectic manifolds \cite{DH}.
Nonabelian generalization was given by Witten in \cite{Witten1}. A
more rigorous treatment is given in \cite{KJ}.

Our analysis of localization in SUSY quantum mechanics will
proceed as follows.

As a first step we show that the field content and symmetries of
the theory allow us to embed the theory into an equivariant
complex. At this point we encounter an obstruction.
 It seems that a bosonic field is missing from the theory. The
cure is to add one extra field, modify the symmetry of the theory
by making it act on the new variable nontrivially and do all this
without changing the action. As a result of this construction we
obtain a modified fermionic symmetry and a new bosonic symmetry
with the crucial property that the latter is the square of the
former. This allows us to interpret the fermionic symmetry
operator as an equivariant differential operator and thus embed
the theory into an equivariant complex. We note that the extra
bosonic field we introduce is not the usual $D$ term which is
already present in the action. In fact in our construction the $D$
term becomes the parameter of the bosonic symmetry. Next we show
that the action of the theory is an equivariantly closed form.
Using this fact, we apply localization principle to the partition
function and show that the path integral localizes on the
instantons of the action.

In the first few sections of the article we review the basics of
equivariant cohomology and discuss the field theoretic
generalizations. In Sec.4 and the first part of Sec.5, following
\cite{Witten1}, we derive localization formula and show how it can
be applied to localize integrals on symplectic manifolds with
Hamiltonian group action. The remainder of the article is devoted
to the discussion of SUSY quantum mechanics in the context of
equivariant cohomology and localization principle.

\section{Equivariant Cohomology}

Let $M$ be a smooth manifold and G a simple, compact Lie group
with a smooth action on $M$. We will denote the Lie algebra of G
by $\mathcal{G}$. A typical element in $\mathcal{G}$ will be
written as $\phi^{a}T_{a}$ where $T_{a}$'s are the generators of
the Lie algebra. Killing-Cartan form will be denoted by $(\:,\:)$.
The infinitesimal generator of the action corresponding to the
element $\sum_{a}\phi^{a}T_{a}$ of the Lie algebra is
$V(\phi)=\sum_{a}\phi^{a}V_{a}$ where $V_{a}$ is the infinitesimal
generator corresponding to $T_{a}$. From now on we will use
summation convention.

We will also need the algebra of formal power series over
$\mathcal{G}$; it will be denoted by $P(\mathcal{G})$. This
algebra is isomorphic to the symmetric algebra on the dual of the
Lie algebra
\begin{equation}
    P(\mathcal{G})\cong S(\mathcal{G}^{*}).
\end{equation}
As a consequence of this isomorphism G acts on $P(\mathcal{G})$ by
the co-adjoint action whose infinitesimal version is
\begin{equation}
    ad^{*}(\psi^{b}T_{b})\phi^{a}=-\psi^{b}\,c^{a}_{bc}\,\phi^{c}.
\end{equation}
Here $c^{a}_{bc}$ are the structure constants of $\mathcal{G}$.

$\Omega_{G}^{*}(M)$ is defined as the G-invariant part of the
tensor product of the exterior algebra on $M$ with
$P(\mathcal{G})$: $\Omega_{G}^{*}(M)=(\Omega^{*}(M)\otimes
P(\mathcal{G}))_{G}$. A typical element of $\Omega_{G}$ is of the
form $\sum_{k} \alpha^{k} P_{k}(\phi)$ and is invariant under the
group action whose infinitesimal version is
\begin{equation}
    \pounds_{V(\psi)}\otimes 1+1\otimes ad^{*}(\psi^{b}T_{b}).
\end{equation}
Here $\alpha^{k}$ is a k-form on $M$, $P_{k}$ is a polynomial in
the Lie algebra and $\pounds_{V(\psi)}$ is the Lie derivative with
respect to the infinitesimal generator $V(\psi)$. Multiplication
in this algebra is given by
\begin{equation}
    \alpha^{k} P_{k}(\phi)\cdot \beta^{l}
    R_{l}(\phi)=(\alpha^{k}\wedge\beta^{l})P_{k}(\phi)R_{l}(\phi).
\end{equation}

Alternatively one can represent elements of $\Omega_{G}^{*}(M)$ as
polynomials in $\mathcal{G}$ with differential forms as
coefficients
\begin{equation}\label{}
    \sum_{I}\sum_{k} \alpha_{I}^{k} \phi^{I}.
\end{equation}
Here $I$ is a multi-index, $\alpha_{I}^{k}$ a k-form on $M$ and
$\phi^{I}$ a monomial in $P(\mathcal{G})$. This representation
allows us to define a grading on $(\Omega_{G}^{*}(M)$ by declaring
the degree of $\alpha_{I}^{k} \phi^{I}$ to be $k+2\mid{I}\mid$.
\begin{equation}\label{}
\Omega_{G}^{*}(M)=\bigoplus_{i\geq 0}\Omega_{G}^{i}(M)
\end{equation}
where
\begin{equation}
    \Omega_{G}^{i}(M)=span\{\alpha_{I}^{k} \phi^{I}\in\Omega_{G}^{*}(M)
    :i=k+2\mid{I}\mid\}.
\end{equation}

Now let us consider the operator
\begin{equation}
  D_{\phi}=d-i\phi^{a}\iota_{V_{a}}
\end{equation}
Here $\iota$ is the contraction. The action on $\Omega_{G}^{*}(M)$
is given by
\begin{equation}
      D_{\phi}\sum_{I}\sum_{k} \alpha_{I}^{k} \phi^{I}=[ d
      \alpha_{I}^{k}+i\phi^{a}\iota_{V_{a}}\alpha_{I}^{k}]\phi^{I}.
\end{equation}
Note that
\begin{equation}
    D_{\phi}:\Omega_{G}^{i}\rightarrow \Omega_{G}^{i+1}
\end{equation}
and
\begin{equation}
D_{\phi}^{2}=-i\pounds_{V(\phi)}
\end{equation}
the last statement implies that $D_{\phi}^{2}=0$ on
$\Omega_{G}^{*}(M)$.

It is also easy to show that for $\alpha\in A_{i}$ and $\beta\in
A_{j}$
\begin{equation}
    D_{\phi}(\alpha\beta)=(D_{\phi}\alpha)\beta+(-1)^{i}\alpha D_{\phi}\beta
\end{equation}
therefore D is a graded differential of degree 1.

Now we have all the ingredients needed for a cohomology theory,
namely, a graded algebra and a graded differential operator of
degree 1. The resulting cohomology is called the Cartan model of
equivariant cohomology and is denoted by $H_{G}^{*}(M)$.

\section{Field Theoretic Generalization}

Let us now show how equivariant cohomology arises in quantum field
theories with BRST like symmetries. More precisely we will
consider field theories which involve three types of fields:
dynamical bosonic fields $A_{a}$, anticommuting ghosts $\psi^{a}$
and auxiliary bosonic fields $\phi^{a}$. We will require the
theory to posses two symmetries; one bosonic and one fermionic.
The bosonic symmetry, which from now on will be referred to as
gauge symmetry, will effect only $A_{a}$'s by mixing them with
$\phi^{a}$'s. The fermionic symmetry will be called BRST symmetry
and the infinitesimal action will be denoted by $\delta_{BRST}$.
The crucial point is that these two symmetries are related to each
other by the following relations

\begin{eqnarray}
   \delta_{BRST}A^{a}&=& \psi^{a}  \\
   \delta_{BRST}\psi^{a}&=& -i\delta_{gauge}A^{a}  \\
   \delta_{BRST}\phi^{a}&=&0
\end{eqnarray}
which implies
\begin{equation}
       \delta_{BRST}^{2}=-i\delta_{gauge}.
\end{equation}

Roughly speaking one can define the BRST symmetry as the square
root of the gauge symmetry. Note that it is the inclusion of the
ghosts to the theory which makes it possible to take the square
root of the gauge transformation. Incidentally this is very
similar to the definition of imaginary numbers as square roots of
negative numbers. In order to take square roots of negative
numbers we have to enlarge $\mathbf{R}$ to
$\mathbf{C}=\mathbf{R}\oplus \mathbf{R}$.

A prototypical example of a BRST like theory is of course the
gauge fixed Yang-Mills theory in the first order formalism where
$A_{a}$'s are the gauge fields, $\psi^{a}$'s are the Faddeev-Popov
ghosts and $\phi^{a}$'s are the Lie algebra valued auxiliary
fields.

Note that the BRST operator is nilpotent (of order 2) on the gauge
invariant sector of the theory. This is again the familiar
situation encountered in gauge theories.

In order to see the relevance of equivariant cohomology to field
theory let us go back to the finite dimensional case and consider
the action of the equivariant differential on the generators of
$\Omega_{G}(M)$
\begin{eqnarray}
   D_{\phi}x^{i}&=&dx^{i}  \\
   D_{\phi}dx^{i}&=&-i\phi^{a}V_{a}^{i}  \\
   D_{\phi}\phi^{a}&=&0.
\end{eqnarray}
Comparing this with the relation between the gauge and BRST
transformations, and using the following identifications in
passing from finite dimensional case to field theory
\begin{eqnarray}
   x^{i}&\rightarrow& A^{a}  \\
   dx^{i}&\rightarrow& \psi^{a}  \\
   \phi^{a}&\rightarrow& \phi^{a}
\end{eqnarray}
we conclude that $\delta_{BRST}$ and $\delta_{gauge}$ are the
field theory analogs of $D_{\phi}$ and $\pounds_{V(\phi)}$
respectively.

Note that this identification also implies the interpretation of
ghost fields as infinite dimensional differential forms.

Thus we see that our BRST like field theory provides us with an
equivariant cohomology where the BRST operator plays the role of
the Cartan differential $D_{\phi}$.

\section{Equivariant Localization}

Let us go back to the finite dimensional case and see how one can
integrate equivariant forms. The field theoretic generalization
will involve path integrals.

The integral of an equivariant form is defined as
\begin{equation}
    \oint\sum_{k}\alpha^{k}P_{k}(\phi)=\sum_{k}\int_{M}\alpha^{k}\int[d\phi]P_{k}(\phi).
\end{equation}
Here the $\phi$ integration is over all $\mathcal{G}$ and
$[d\phi]$ is a measure on the Lie algebra which includes a
Gaussian convergence factor
\begin{equation}
         [d\phi]=\frac{d\phi^{1}\ldots d\phi^{n}}{VolG}\:e^{-\frac{\epsilon}{2}(\phi,\phi)}
\end{equation}
A very useful property of this integral is that if $M$ is without
boundary then the integral of an equivariantly exact form
vanishes. In other words integration by parts produces a vanishing
surface term
\begin{equation}
    \oint_{M}D\alpha=0
\end{equation}
provided $\partial{M}$ is empty.

\paragraph{Localization Principle.}
Let $\alpha$ be an equivariantly closed form and $\lambda$ an
equivariant 1-form. Then it is easy to show that the following
form is equivariantly exact
\begin{equation}
    \alpha(1-e^{tD_{\phi}\lambda}).
\end{equation}
Integrating this exact form over M we obtain the very important
formula which forms the basis of equivariant localization
\begin{equation}
    \oint\alpha=\oint\alpha e^{tD_{\phi}\lambda}.
\end{equation}
So the integral on the right hand side is in fact independent of
the value of the parameter t and is equal to the integral of
$\alpha$. In particular $\oint\alpha$ can be calculated by
evaluating $\oint\alpha e^{tD_{\phi}\lambda}$ in the asymptotic
limit $t\rightarrow\infty$. So let us take a closer look at this
integral
\begin{equation}
    \oint\alpha e^{tD_{\phi}\lambda}=\oint\alpha
    e^{td\lambda+i\sum_{a}\phi^{a}\lambda(V_{a})}.
\end{equation}
Since the two terms appearing in the argument of the exponential
commute with each other we can separate the exponential as
$e^{td\lambda}\,e^{i\sum_{a}\phi^{a}\lambda(V_{a})}$. The
contribution of the first factor to the integral is just a
polynomial in $t$. Therefore in the limit $t\rightarrow\infty$ the
integral is localized on the critical points of the function
$\sum_{a}\phi^{a}\lambda(V_{a})$:
\begin{equation}
    \lambda(V_{a})=0
\end{equation}
and
\begin{equation}
    \phi^{a}d\lambda(V_{a})=0.
\end{equation}
From now on we will assume that the only solution of the last
equation is $\phi^{a}=0$ for all $a$.

\section{Localization on Symplectic Manifolds}

In this section we want to localize the integral of a specific
equivariantly closed form on a symplectic manifold ($M$,$\omega$).
Here $\omega$ is the symplectic form on $M$. We assume that a
compact Lie group G with a simple Lie algebra $\mathcal{G}$ acts
on $M$. Moreover, the action will be assumed to be symplectic and
Hamiltonian. The moment map $\mu_{a}$ corresponding to the action
of the generator $T_{a}$ of $\mathcal{G}$ is given by
\begin{equation}
    d\mu_{a}=-\iota_{V_{a}}\omega.
\end{equation}
We will also define
\begin{equation}
    I=\sum_{a}\mu_{a}^{2}.
\end{equation}
The equivariant form that want to integrate is $e^{\bar{\omega}}$
where
\begin{equation}
    \bar{\omega}=\omega-i\phi^{a}\mu_{a}.
\end{equation}
is the so called equivariant symplectic form. An easy calculation
\cite[Chapter 9]{GS} shows that $\bar{\omega}$ is G-invariant and
equivariantly closed.

The explicit form of the integral in question is
\begin{equation}
    \int_{M}\int \frac{d\phi^{1}\ldots
    d\phi^{n}}{VolG}\:\exp\left[{\omega-i\phi^{a}\mu_{a}-\frac{\epsilon}{2}(\phi,\phi)}\right]
\end{equation}
In order to apply localization to this integral we have to choose
a convenient $\lambda$. This can be done if we assume that there
exists a G-invariant almost complex structure $J$ on $M$ whose
associated metric $g(X,Y)=\omega(JX,Y)$ is positive definite. Then
\begin{equation}
    \lambda=JdI
\end{equation}
is a suitable choice.

It is shown in \cite{Witten1} that the condition
$\lambda(V_{a})=0$ for all $a$ is equivalent to $dI=0$. Thus the
integral in question localizes on the critical set of the function
I. Writing down the condition $dI=0$ explicitly
\begin{equation}
    \sum_{a}\mu_{a}d\mu_{a}=0
\end{equation}
we see that there are two types of critical points: those coming
from the simultaneous vanishing of moment maps and those that
don't (higher critical points). In what follows we will ignore the
higher critical points.

\paragraph{Example.} Let us consider $\mathbf{R^{2n}}$ with a
symplectic form
\begin{equation}
    \omega=\omega_{ij}(q)dp^{i}\wedge dq^{j}.
\end{equation}
Here the matrix $\omega_{ij}(q)$ is a function of q coordinates
only. The closedness of $\omega$ implies
\begin{equation}
     \frac{\partial \omega_{jk}}{\partial q^{i}}-\frac{\partial \omega_{ji}}{\partial q^{k}}=0
\end{equation}
Now the action of the additive group $\mathbf{R^{n}}$ given by
\begin{eqnarray}
    q^{i} & \rightarrow & q^{i} \\
  p^{j}   & \rightarrow & p^{j}+a^{j}
\end{eqnarray}
is symplectic and Hamiltonian. Infinitesimal generators of this
action are
\begin{equation}
    V_{k}=\frac{\partial}{\partial p^{k}}
\end{equation}
The moments are given implicitly by the solutions of
\begin{eqnarray}
  \frac{\partial \mu_{k}}{\partial q^{i}}&=& -\omega_{ki} \\
   \frac{\partial \mu_{k}}{\partial p^{i}}&=& 0.
\end{eqnarray}
In this example we have relaxed the compactness condition on the
group $G=\mathbf{R^{n}}$. This is not harmful in the case of
$\mathbf{R^{n}}$ which admits a bi-invariant measure anyway. In
any case, it is a simple matter to put periodicity conditions on
the coordinates and replace $\mathbf{R^{n}}$ by an appropriate
toric group.

\section{Localization for SUSY Quantum Mechanics}

Here we have a 0+1 dimensional theory on the circle with the
action
\begin{equation}
    S=\int dt \left[\frac{dx}{dt}+\frac{\partial V(x)}{\partial x
    }\right] (iD)+\frac{1}{2}D^{2}-i\overline{\psi}\left[\frac{d}{dt}+\frac{\partial^{2}V(x)}{\partial
    x^{2}}\right]\psi.
\end{equation}
This theory is invariant under a SUSY transformation of the form
\begin{eqnarray}
  \delta_{SUSY}x &=& \psi \\
  \delta_{SUSY}\psi &=& 0 \\
  \delta_{SUSY} \overline{\psi} &=& -i (iD) \\
  \delta_{SUSY}(iD) &=& 0 \\
  \delta_{SUSY}^{2} &=& 0.
\end{eqnarray}
We see that this SUSY transformation is  nilpotent of degree 2.
However $\delta_{SUSY}^{2}$ vanishes identically and therefore our
theory does not seem to be a BRST like theory as defined above.
The problem is that there is no bosonic gauge symmetry at sight.
The solution is to introduce an extra variable without changing
the theory and extend the SUSY transformation in such a way that
its square gives us a new bosonic symmetry. Thus we introduce a
new bosonic field $\pi$ and replace the SUSY transformation by the
BRST transformation
\begin{eqnarray}
  \delta_{BRST}x &=& \psi \\
  \delta_{BRST}\psi &=& 0 \\
  \delta_{BRST}\pi &=& \overline{\psi} \\
  \delta_{BRST}\overline{\psi} &=& -i (iD) \\
  \delta_{BRST}(iD) &=& 0.
\end{eqnarray}
or
\begin{equation}
 \delta_{BRST}=\int dt \: \psi(t)\,\frac{\delta}{\delta x(t)}+\overline{\psi}(t)\,\frac{\delta}{\delta
 \pi(t)}-i(iD)(t)\,\frac{\delta}{\delta \overline{\psi}(t)}.
\end{equation}
Now we have
\begin{equation}
  \delta_{BRST}^{2}=-i\delta_{gauge}
\end{equation}
where
\begin{eqnarray}
   \delta_{gauge}\pi&=& iD \\
   \delta_{gauge}(x, \psi, \overline{\psi}, iD)&=&0.
\end{eqnarray}
or
\begin{equation}
        \delta_{gauge}=\int dt \: (iD)(t)\frac{\delta}{\delta \pi(t)}
\end{equation}
Note that $\delta_{BRST}$ agrees with $\delta_{SUSY}$ on the
original variables so it is a symmetry of the theory. Moreover
$\delta_{gauge}$ is a symmetry since it effects only the extra
variable $\pi$ which does not even appear in the action. In this
way we turn our theory into a BRST like theory.

Our next task is to look for possible applications of
localization. We will show that the partition function localizes
on the instantons of the theory.

\paragraph{Symplectic Geometry of the Field Space and Localization.}
The bosonic fields of the theory are the coordinates of the field
space $M={(x(t),\pi(t))}$. Ghost fields are interpreted as
differentials of the coordinates $\delta x(t)=\psi(t)$ and $\delta
\pi(t)=\overline{\psi}(t)$.

The (pre)symplectic structure on the field space is defined as
\begin{equation}
     \int dt\:dt'\: \overline{\psi}(t')\, \omega[t',t;x]\,\psi(t)
\end{equation}
where
\begin{equation}
 \omega[t',t;x]=i\left[-\frac{d}{dt}+\frac{\partial^{2}V(x)}{\partial
    x^{2}}\right]\delta(t-t').
\end{equation}
This form is closed since
\begin{eqnarray}
  \frac{\delta \omega[t',t;x]}{\delta x(t'')} &=& V'''(x(t'))\delta(t'-t'')\delta(t'-t) \\
  \frac{\delta \omega[t',t'';x]}{\delta x(t)}&=& V'''(x(t'))\delta(t'-t)\delta(t'-t'')
\end{eqnarray}
\begin{equation}
\frac{\delta \omega[t',t;x]}{\delta x(t'')}-\frac{\delta
\omega[t',t'';x]}{\delta x(t)}=0.
\end{equation}
The action of the gauge transformation is symplectic since
\begin{eqnarray}
  \delta_{gauge}(x, \overline{\psi}, \psi)&=& (0,0,0) \\
   \delta_{gauge}\int dt\, dt'\: \overline{\psi}(t')
   \,\omega[t',t;x]\,\psi(t)&=&0.
\end{eqnarray}
In order to show that the action is Hamiltonian we have to solve
the moment equations
\begin{eqnarray}
   \frac{\delta \mu[t';x]}{\delta x(t)}&=& -\omega[t',t;x] \\
   \frac{\delta \mu[t';x]}{\delta \pi(t)}&=& 0.
\end{eqnarray}
These can be integrated to give
\begin{equation}
    \mu[t';x]=i\left[\frac{dx(t')}{dt'}+V'(x(t'))\right].
\end{equation}
So the action is
\begin{equation}
    S=\overline{\omega}[x]+\int dt \: \frac{1}{2}D^{2}.
\end{equation}
where
\begin{equation}
    \overline{\omega}[x]=\omega[x]-i\int dt \:\mu[t;x](iD)(t)
\end{equation}
is the equivariant symplectic form on the field space.

The partition function is
\begin{equation}
    Z=\int\,  \textrm{D}D\,\textrm{D}x\, \textrm{D}\psi\,
    \textrm{D}\overline{\psi}\:
    \exp i\left[\omega[x]-i\int dt\mu[t;x](iD)(t)+\int dt \frac{1}{2}D^{2}\right].
\end{equation}
This is almost in the form of equation (34). All we need to do to
arrive at the correct form (up to a harmless factor of i) is to
integrate over the new variable $\pi$ and divide by the volume of
the gauge group. Since $VolG=\int\,\textrm{D}\pi$ these operations
do not change the value of the partition function. So we conclude
that the partition function localizes on the zeroes of the moment
map, that is, the instantons of the theory
\begin{equation}
      \left[\frac{dx(t)}{dt}+V'(x(t))\right]=0.
\end{equation}
Integrating the square of this expression and discarding the
surface term yields:
\begin{eqnarray}
   \frac{dx(t)}{dt}&=& 0  \\
    V'(x(t)) &=& 0,
\end{eqnarray}
so localization is on the critical points of the potential $V$.

\paragraph{Acknowledgement.} I would like to thank O.T. Turgut
for helpful discussions.

\bibliographystyle{plain}

\end{document}